\documentclass{aa}

\usepackage{graphicx}
\usepackage{mathabx}
\usepackage{color}
\usepackage{txfonts}
\begin{document} 

   \title{Volatile-rich comets ejected early on during Solar System formation}

   \author{S.E. Anderson \inst{1}, J.-M. Petit\inst{1}, B. Noyelles\inst{1}, O. Mousis\inst{2}, \and P. Rousselot\inst{1}
          }

   \institute{Institut UTINAM, UMR 6213 CNRS - Univ. Bourgogne Franche-Comt\'e, OSU THETA, BP 1615, F-25010 Besançon Cedex, France\\
              \email{sarah.anderson@univ-fcomte.fr}
         \and
             Aix Marseille Univ, Institut Origines, CNRS, CNES, LAM, Marseille, France\\
             }

   \date{Accepted September 8, 2022}
   \authorrunning{S. E. Anderson et al.}
 
  \abstract
   {Comet C/2016 R2 PanSTARRS (hereafter C/2016 R2) presents an unusually high N$_2$/CO abundance ratio, as well as a heavy depletion in H$_2$O, making it the only known comet of its kind. Understanding its dynamical history is therefore of essential importance as it would allow us to gain a clearer understanding of the evolution of planetesimal formation in our Solar System. Two studies have independently estimated the possible origin of this comet from building blocks formed in a peculiar region of the protoplanetary disk, near the ice line of CO and N$_2$.}
   {We intend to investigate the fates of objects formed from the building blocks in these regions. We hope to find a possible explanation for the lack of C/2016 R2-like comets in our Solar System.}
   {Using a numerical simulation of the early stages of Solar System formation, we track the dynamics of these objects in the Jumping Neptune scenario based on five different initial conditions for the protosolar disk. We integrate the positions of 250\,000 planetesimals over time in order to analyze the evolution of their orbits and create a statistical profile of their expected permanent orbit.}
   {We find that objects formed in the region of the CO- and N$_2$- ice lines are highly likely to be sent towards the Oort Cloud or possibly ejected from the Solar System altogether on a relatively short timescale. In all our simulations, over 90\% of clones formed in this region evolved into a hyperbolic trajectory, and between 1\% and 10\% were potentially captured by the Oort Cloud. The handful of comets that remained were either on long-period, highly eccentric orbits like C/2016 R2, or absorbed into the Edgeworth–Kuiper belt.}
   {Comets formed <15~au were predominantly ejected early in the formation timeline. As this is the formation zone likely to produce comets of this composition, this process could explain the lack of similar comets observed in the Solar System.}


   \maketitle
%

\section{Introduction}

Comets are some of the most pristine bodies in the Solar System, having remained relatively unchanged since their formation 4.6 billion years ago. Cometary nuclei provide insights into the composition of the early protoplanetary disk (PPD) through their isotopic abundance ratios. As their composition reflects the physico-chemical conditions of the disk at the location of their formation in the protosolar nebula (PSN), understanding where each comet was formed reveals details as to the evolution of the Solar System.

Decades of remote sensing of comets have revealed these objects to be water-ice rich, with a typical carbon monoxide composition of CO/H$_2$O = 4\% \citep{Bockelee2017}, and depleted in N$_2$ despite the abundance of this molecule in the atmospheres and surfaces of the outer Solar System bodies, such as Triton or Pluto \citep{Cochran2000}. However, radio observations of the long-period comet C/2016 R2 (PanSTARRS) revealed that its composition is unlike any comet observed before, with the spectrum dominated by bands of CO$^+$. This CO-rich comet is remarkably depleted in water, with a H$_2$O/CO ratio of only $\sim0.32$ \%  \citep{McKay2019} with an upper limit of H$_2$O/CO < 0.1 \citep{Biver2018}. Further, it has a peculiar abundance of N$_2^+$, with N$_2$/CO estimated to be between 0.05 $\pm$ 0.01 \citep{McKay2019}, 0.06 $\pm$ 0.01 \citep{Opitom2019}, and 0.08 $\pm$ 0.02 \citep{Biver2018}, which had never been seen in such high quantities in comets before. This composition changes our perception of comet formation, as it was previously understood that CO ice is unlikely freeze out without abundant water ice, which has a higher binding energy than CO \citep{Boogert2015}. Most volatile species would also be expected to deplete with each subsequent passage of this comet within the inner Solar System. Understanding the dynamical history of this comet is therefore of essential importance to understanding the timeline of planetesimal formation in our Solar System. 

Other potential N$_2$-presenting candidates have been identified, such as C/1908 R1 (Morehouse), C/1961 R1 (Humason), C/1987 P1 (Bradfield), C/2001 Q4 (NEAT) with N$_2$/CO=0.027 \citep{Feldman2015}, and C/2002 VQ94 (LINEAR) for which N$_2$/CO=0.06 \citep{Korsun2008}. A few short-period comets also show an increased N$_2$/CO ratio, such as comet 29P/Schwassmann-Wachmann 1 with N$_2$/CO=0.013 \citep{Ivanova2016}, or comet 67P presenting N$_2$/CO=0.0287, but this result came from in situ measurements \citep{Rubin2020}. Some others present moderately unusual water-poor compositions; for example, interstellar comet 2I/Borisov, measured to have CO/H$_2$O between 35\% and 173\% \citep{Cordiner2020, Bodewits2020}, which is significantly higher than the average cometary values for our Solar System, and could be explained by an unusual formation environment beyond the CO snow line of its own system \citep{Price2021}. Comet C/2009 P1 (Garradd) is another outlier with a CO production rate of 63\%  of that of water, yet no N$_2$ was detected \citep{Feaga2014}. This simultaneously CO- and N$_2$-rich and water-poor composition, along with none of the usual neutrals seen in most cometary spectra, makes C/2016 R2 a unique and intriguing specimen, the only one of its kind to ever be observed.

Such a small sample size makes it impossible to draw conclusions as to a shared formation reservoir. The long-period comets share highly eccentric, almost parabolic orbits ---even hyperbolic in the case of C/1908 R1 (Morehouse) and C/2001 Q4 (NEAT)---, while Comet 29P/Schwassmann-Wachmann 1 is likely a captured Oort Cloud object \citep{Neslusan2017}. It is clear these objects must have spent the majority of their lifetime at high heliocentric distance, else they would have already lost their volatile content. Unfortunately, attempts to trace back their dynamical history with any degree of certainty is made impossible by the inherent chaotic nature of their motion due to frequent close encounters with the gas giants, which strip comets of their dynamical memory. As a result, there is no peculiar C/2016 R2-like orbit despite its otherwise peculiar nature, and we cannot trace its dynamical history backwards to a potential shared formation reservoir.

The origin of the unusual composition of C/2016 R2 is highly disputed. It may be a fragment of a differentiated object as suggested by \cite{Biver2018}, similar to the CO-rich interstellar comet 2I/Borisov \citep{Cordiner2020}. If CO is absent in the upper layers of an as-of-yet undiscovered differentiated comet, as suggested by \cite{DeSanctis2001}, then it is possible that C/2016 R2 is a fragment of the core of such a comet. \citet{Desch2021} theorize that 1I/'Oumuamua may be an N$_2$ iceberg chipped off from the surface of an ex-Pluto by an impact during a period of dynamical instability, which could be applied to C/2016 R2. Another possibility is that the particular
composition of C/2016 R2  simply arises from where it formed in the PSN: Perhaps this disk could evolve over time to create exotic compositions at different disk locations in unique proportions, in "special" comet-forming annuli. Two studies independently estimated the possible origin of this comet from building blocks formed in a peculiar region of the PSN, near the ice lines of CO and N$_2$. By evaluating the radial transport of volatiles in the PSN, \cite{Mousis2021} found that the  peculiar N$_2$/CO ratio of  C/2016 R2 could be replicated by agglomeration from particles near the N$_2$ and CO ice lines, within the 10-15 au region. Meanwhile, the CO/H$_2$O ratio would remain deeply depleted inward of the CO ice line, around the 8-11 au region. Cold traps of hypervolatiles in the PSN in a small, specific region of the disk could explain the peculiar composition of this latter comet. Similarly, \cite{Price2021} model the effect of drifting solid material in the PPD and find that the ideal location for the formation of CO-rich, H$_2$O-poor objects is beyond CO ice line. However, this would seem to indicate that more CO-rich comets should exist than have previously been observed. The N$_2$/CO ratio was not a part of their study. 

Here we explore the potential fates of comets formed from these building blocks using a numerical simulation of early Solar System formation. By examining the dynamical evolution of only the objects formed in a small exotic pocket, or "Sweet Spot," of the PSN, which allows for peculiar-composition comets to form, we hope to understand why so few are observed today. In Section \ref{sec:meth}, we describe the model we use to simulate the early Solar System and the dynamical evolution of these small bodies. In Section \ref{sec:res}, we report on these results and examine more closely the fates of all comets, then narrow our interest to comets that would have formed inside the Sweet Spot. Finally, in Section \ref{sec:con}, we provide our conclusions as to what these fates will be.

\section{Methods}\label{sec:meth}

We employ the Jumping Neptune scenario from \cite{Nesvorn2015}. We begin with five planets: Jupiter, Saturn, and three ice giants of comparable mass, as described by \cite{Deienno2017}. This third ice giant, henceforth I1, undergoes a series of encounters with Jupiter and Saturn which causes a divergent jump in their semi-major axes before inducing a jump in Neptune's orbit as well. Finally, I1 is ejected onto a hyperbolic orbit, leaving the remaining four planets near their present-day orbits. We tested several alternative simulations, varying the multi-resonance configuration, the distance from the last planet to the inner edge of the disk (1 or 2 au), the mass of the disk (20 or 40 $M_\Earth$), and the inclination of the disk in relation to the plane of the planets, with five different evolutions in each case. We selected the simulations that best satisfy the criteria of similarity with the Solar System today, consistent with the current orbital structure of the trans-Neptunian population, in line with \citet{Deienno2017}, which were all from the 3:2, 3:2, 2:1, 3:2 multi-resonance configuration, with a disk of 40 $M_\Earth$. The initial multi-resonant configurations we choose for Jupiter, Saturn, Uranus, Neptune, and I1, along with parameters for the location and mass of disk, begin in a 3:2, 3:2, 2:1, 3:2 resonance, as \cite{Baguet2019} find this is able to place a secular tilt resonance in the area of the cold Edgeworth-Kuiper belt (between 39 and 48 au). This provides us with five scenarios to explore, as defined in Table \ref{table:1}. All these configurations require the existence of the fifth giant planet, with a mass comparable to those of Uranus or Neptune, which is eventually ejected during the instability. This ice planet would have formed within the volatile-rich zone identified by \citet{Mousis2021}. All the planetary evolution simulations were run self-consistently with the five planets and a swarm of 1000 massive particles of the same mass, each 1/1000th of the mass of the disk. The disk extends from its inner edge to 30 au.

\begin{figure}
    \centering
    \includegraphics[width=0.48\textwidth]{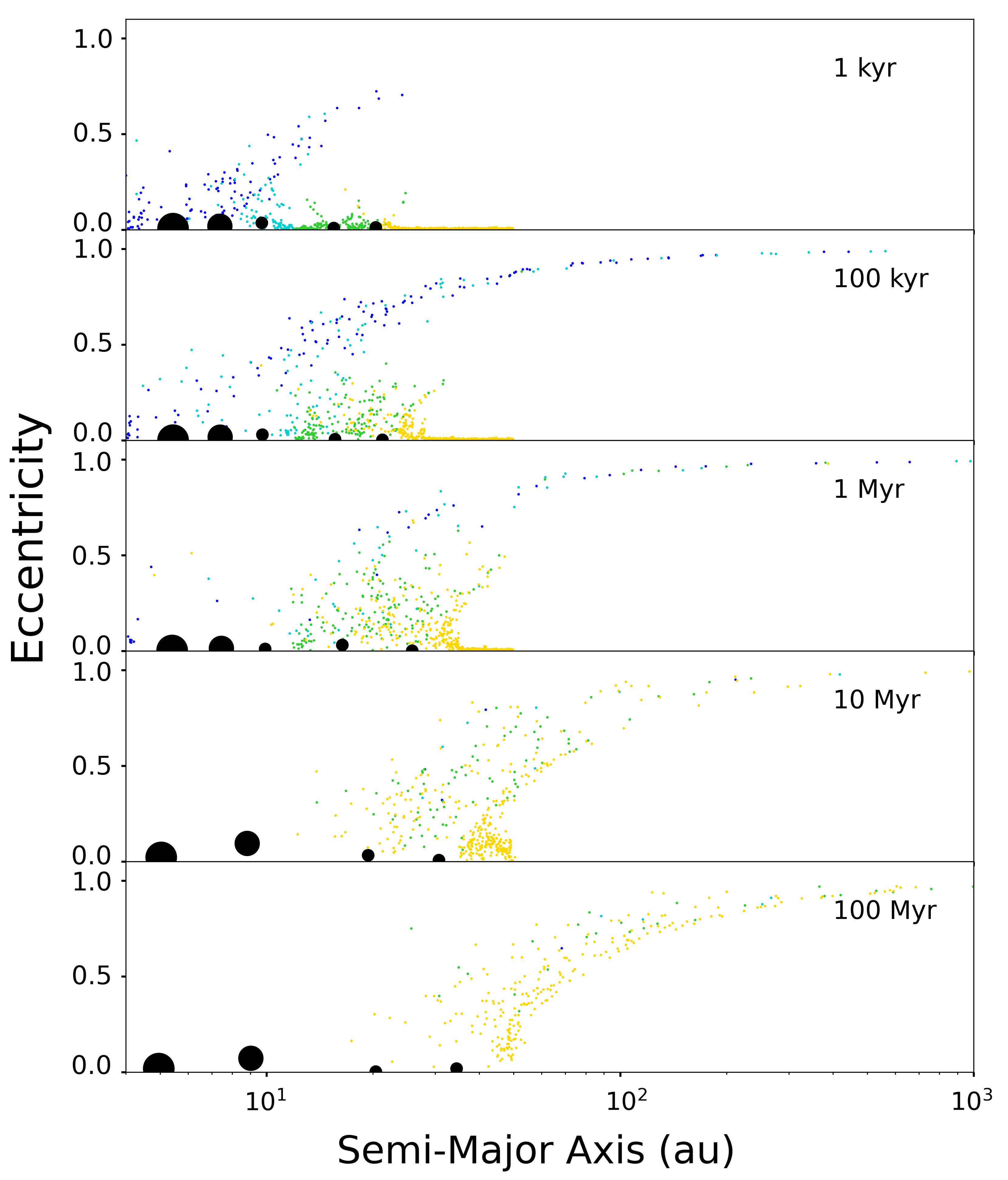}
    \caption{Dynamical evolution of the planets and 1000 comet clones in Scenario 1 with eccentricity as a function of semi-major axis in log scale. Planets are represented in black, with Jupiter, Saturn, I1, Uranus, and Neptune, respectively, from left to right. Blue indicates comets formed between 4 and 8 au. Turquoise indicates comets formed between 8 and 12 au. Green indicates comets formed between 12 and 20 au. Yellow indicates comets formed between 20 and 50 au. After 100 Myr, the area around the giant planets is entirely cleared. Comets formed between 4 and 12 au are the first to be lost, and by 10 Myr almost none remain. By 100 Myr, the comets that remain in our simulation are almost entirely from the 20-50 au population.}
    \label{fig:stacked}
\end{figure}

\begin{figure*}
\sidecaption
\includegraphics[width=12cm]{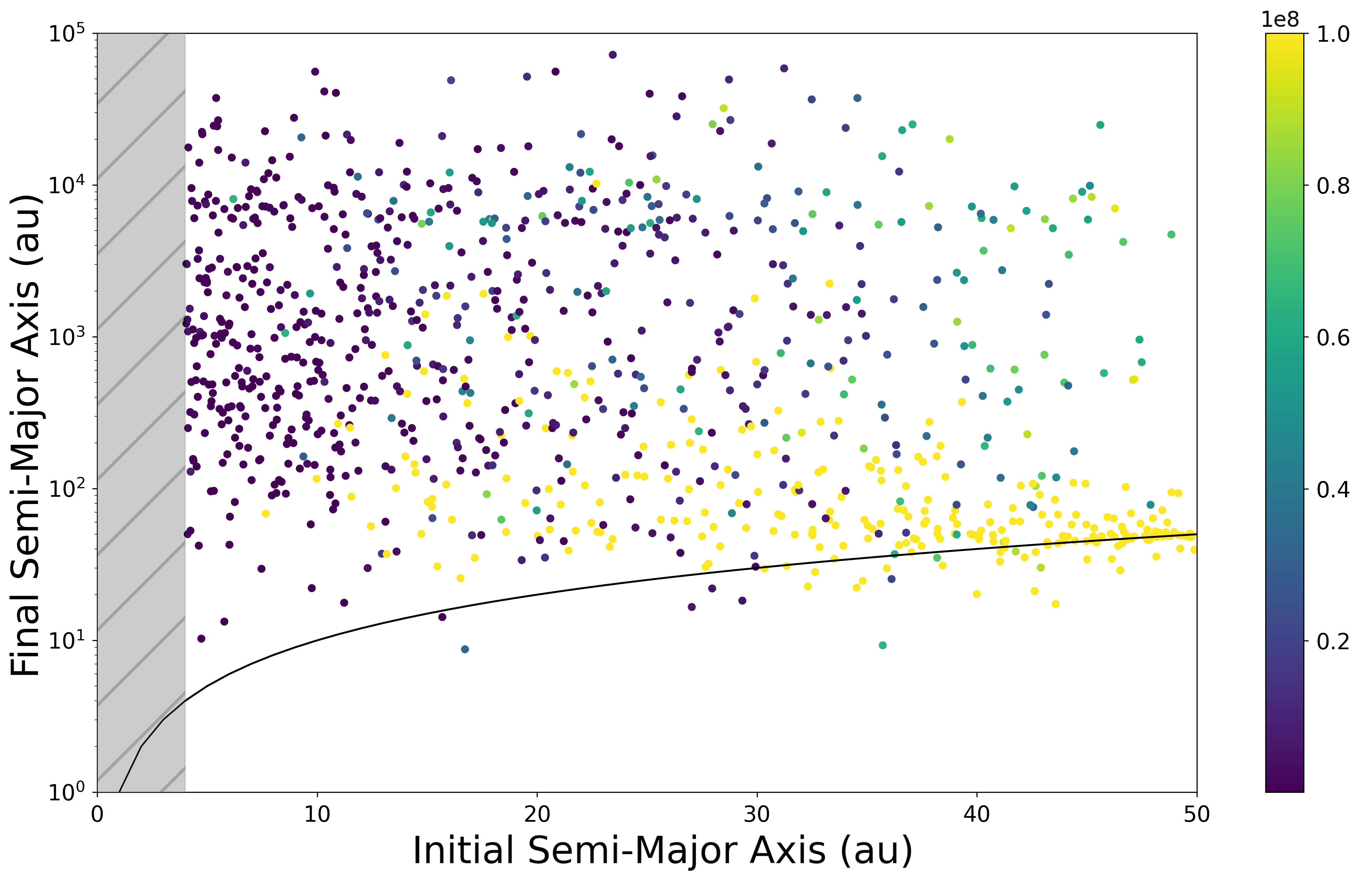}
\caption{Final positions in log scale of the first 1000 clones from Scenario 1 color coded for the last recorded time before they are removed from the simulation (see main text). Violet clones were lost early in the simulation, while yellow clones indicate those that remain. The black line corresponds to an equal initial and final position: comets above would have moved away from their initial formation position, and comets below would have migrated inwards. The area near the orbits of Jupiter and Saturn clears quickly, sending the clones on highly eccentric orbits even before Neptune's migration occurs. We see the relatively stable location of the current day Edgeworth–Kuiper Belt.}
\label{fig:Final}
\end{figure*}

\begin{table}
\caption[]{Initial conditions for the five scenarios explored in this study. The multi-resonance configuration is the 3:2, 3:2, 2:1, 3:2, with the outermost planet at 20.18 au from the Sun. In all cases studied here, we used a disk of 40 $M_\Earth$. (a) Distance of the inner bound of the disk (au), (b) inclination of the disk with respect to the invariable plane ($^\circ$), (c) node of the disk ($^\circ$), and (d) running number, i.e., number of generations used.}          
\label{table:1}      
\centering                          
\begin{tabular}{l c c c c}        
\hline\hline                 
 & (a) & (b) & (c) & (d) \\    
\hline                        
Scenario 1 & 1 & 1 & 100.2 & 08\\
Scenario 2 & 1 & 0 & 0 & 02 \\
Scenario 3 & 2 & 0 & 0 & 04\\
Scenario 4 & 2 & 1 & 100.2 & 01\\
Scenario 5 & 2 & 1 & 100.2 & 05\\
\hline                                   
\end{tabular}
\end{table}

The present-day Edgeworth–Kuiper belt extends from the orbit of Neptune at 30 au to approximately 50 au from the Sun. However, most of the small bodies of the outer Solar System originated from the region between Jupiter and $\sim$30~au \citep{Gomes2003,2005Natur.435..459T,2008Icar..196..258L,Kaib2008}.
With this in mind, we limit our simulations to planetesimals formed in the 4 - 50 au range. This allows us to neglect the influence of the inner planets, which, having small orbits, require more integration steps and longer calculations on each of our clones. While the CO-rich comet-forming zone could extend to 100 au \citep{Price2021}, the mass depletion of the classical belt is already well explored.

We then run a modified \texttt{SWIFT} numerical integrator which uses a pre-recorded evolution of the giant planets \citep{Petit1999} and evolves our system over 100 Myr. The previously calculated evolution of the planets is recorded every 1000 yr or less and the positions of the planets are interpolated at each time-step necessary for the integration of the motion of the test particles. This ensures that each simulation for a given planetary evolution will use exactly the same planetary evolution track, avoiding divergence due to the intrinsic chaotic nature of planetary motion. Thus, our final planetary system is sure to correctly reproduce the structure of the Solar System. The major difference in planetary behavior between these scenarios is the moment of ejection of 
I1. This occurs at 5 Myr, 6 Myr, 7 Myr, 8 Myr, and 13 Myr for scenario 1, 2, 3, 4, and 5, respectively.

For each scenario, we run 50 sets of 1000 massless comet facsimiles or "clones." Each clone has randomly generated orbital elements setting them on the same plane as the disk with varying semi-major axes between 4 au ---to avoid the inner Solar System--- and 50 au. The clones are distributed with a number density that varies as $r^{-1/2}$, or a surface density that varies as $r^{-3/2}$. We therefore have a total of 250\,000 clones for our five scenarios. Our simulations count a clone as lost if it reaches beyond 10000 au as we do not yet have the ability to estimate the effects of the Galactic tidal forces. If a clone moves under 0.005 au from the Sun, or in collision with a planet, it is also removed from the integration, as it is most likely destroyed. 

\section{Results and Discussion}\label{sec:res}

We examine the orbital elements of each clone, identified by its formation location (initial semi-major axis). This is shown in Fig. \ref{fig:stacked}. Within the first 1 Myr, 21\% of all clones are lost from the simulation. This number rises to nearly half (49\%) after 10 Myr. By the end of the 100 Myr simulation run, we have lost three quarters (76\%) of our initial population. Only a quarter (24\%) of our clones remain. A snapshot of the first 1000 clones in our first scenario is shown in Fig.~\ref{fig:Final} with their first and final positions. Each clone is color coded for the moment it is lost, with earlier losses in purple and those that remain in the end shown in yellow. The major loss of clones occurs before $\sim$10 Myr: after this time, the area around the giant planets (<~15 au) is entirely cleared. It is important to note here that if we had used a four-planet model, based on current planetary orbits, Saturn would play the role of I1 and clear this region, leading to the same outcome.

\begin{figure}
   \centering
   \includegraphics[width=\linewidth]{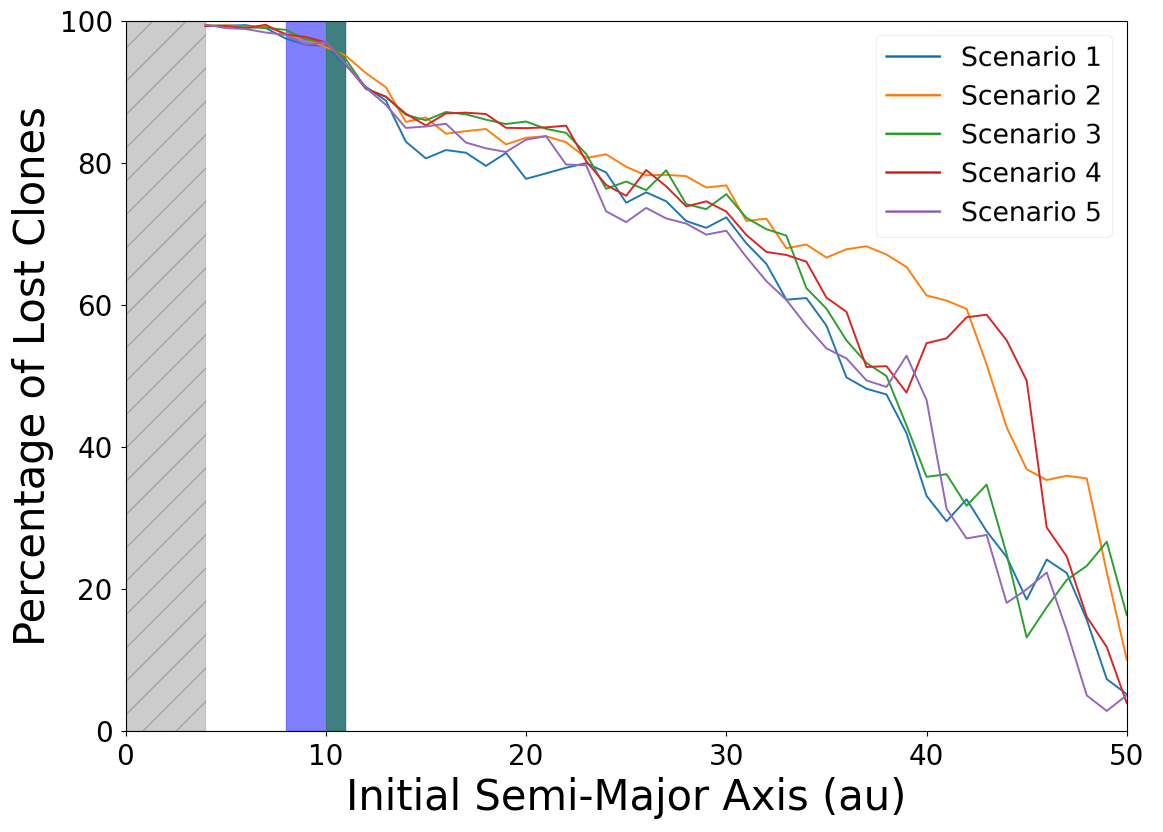}
   \caption{Percentage of clones lost per formation location for each of the five scenarios. The gray zone indicates the limitation of our simulation. The blue zone indicates the N$_2$/CO enrichment zone as predicted by \cite{Mousis2021}, while the overlaid green zone indicates the location of the ideal CO/H$_2$O enrichment zone.}
   \label{fig:PerDis}
\end{figure}

We examine the percentage of clones lost in our simulations  more closely for each 1 au annulus from 4 au to 50 au, as shown in Fig.~\ref{fig:PerDis}. We see that for every 1 au annulus between 4 and 10 au, over 95\% of the clones are lost before the end of the 100 Myr in each scenario. This number dips to 90\% around 12 au. Then, between 12 and 20 au, each scenario still loses a minimum of 80\% of their clones within the simulation time. In comparison, annuli beyond 40 au \textemdash the current location of the Classical Edgeworth–Kuiper Belt\textemdash\  only lose half their clones, showing a zone that is relatively stable, containing objects that do not move far from where they are formed. The behavior of these clones is consistent between scenarios and independent of the moment of ejection of I1.

Based on the ranges proposed by \cite{Price2021} and \cite{Mousis2021}, we examine different formation zones. \cite{Price2021} suggest a wide range, arguing that the CO/H$_2$O ice-enrichment zone is likely between 20 and 100 au, though these authors do not investigate a N$_2$/CO ice-enrichment zone. As the CO/H$_2$O ice enrichment zone evolves over time, and without seeing how N$_2$ would evolve in the simulations of \cite{Price2021}, we cannot determine where a specific C/2016 R2 formation zone could occur. In light of the fact that both CO and N$_2$ have similar sublimation temperatures, the two ice lines should be near each other and make the 20-30 au annulus an area to explore. Meanwhile, the results of \cite{Mousis2021} would indicate a narrow area, as they find a CO/H$_2$O ice-enrichment zone of $\sim$1-2 au wide, near 10 au. Their N$_2$/CO ice-enrichment zone is narrower still, seemingly less than 1 au. The overlapping formation zone for a C/2016 R2-like comet would therefore be incredibly narrow. We examine both a wide C/2016 R2 forming annulus between 8 and 20 au; a narrow one, only 8-10 au; and the narrowest one between 10 and 11 au, that is, the Sweet Spot. Interestingly, I1 is initialized and subsequently ejected from this narrow region as well.

\begin{figure*}
   \centering
   \includegraphics[width=\textwidth]{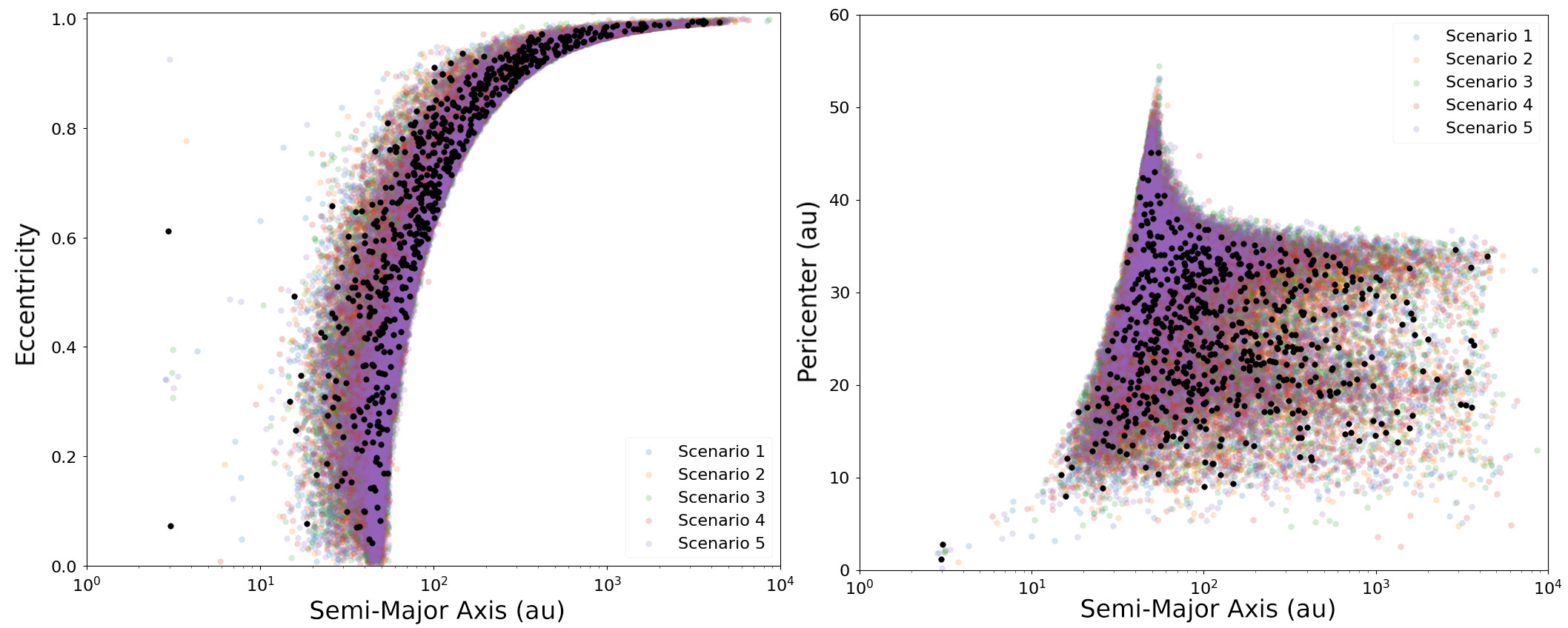}
   \caption{Final semi-major axes and eccentricities (left) and perihelion distances (right) of all clones from all simulations remaining after 100 Myr. Comets formed in the 8-11 au zone are shown in black. Any remaining object with a perihelion $q<35$ au will likely be sent to the Oort Cloud by Neptune, or will lose its hypervolatile majority ices to vacuum via insolation heating \citet{Lisse2022}.}
   \label{fig:RemQ}
\end{figure*}

The resulting statistics are shown in Table \ref{table:2}. On average, each simulation loses 75\% of its clones by 100 Myr, losing 90\% of all clones formed in the 8-20 au range, 97\% of all clones formed in the 8-10 au range, and $\sim80\%$ in the 20-30 range. Consistently, each simulation loses 96\% of all comets formed between 10 and 11 au. If we examine the region of clones initialized between 8 and 20 au, we find that half the clones are already lost by 5 Myr, with two-thirds of clones being ejected after 15 Myr. If we narrow that region further to 8-11 au, we find that 60\% of clones formed in this region are ejected in the first 1 Myr and 90\% after 10 Myr. A handful of clones ($\sim$0.1\%) are lost to collisions with the giant planets. Depending on the chronology, these could help account for the delivery of the building blocks of the Galilean and Saturnian satellites necessary for their formation \citep{Ronnet2018, Anderson2021}. In each simulation, irrespective of the scenario, only $\sim$1\% of all remaining clones were from the initial 8-11 au population; these are shown in black in Fig.~\ref{fig:RemQ}. They will  either find themselves on highly eccentric orbits, be absorbed into the Edgeworth–Kuiper belt,  or join the scattered disk. These clones seem to be evenly distributed within the population of remaining comets. The 10-11 au population makes up only 0.4\% of all surviving clones. 

\begin{table}
\caption[]{Statistical loss outcomes of each of the scenarios after 100 Myr for each formation zone.}            
\label{table:2}      
\centering                          
\begin{tabular}{lccccc}        
\hline\hline                 
 & Total loss & 8-10 au & 10-11 au & 8-20 au & 20-30 au\\    
\hline                        
S1 & 73$\%$ & 97$\%$ & 96$\%$ & 88$\%$ & 76$\%$ \\
S2 & 79$\%$ & 97$\%$ & 96$\%$ & 90$\%$ & 80$\%$ \\
S3 & 75$\%$ & 98$\%$ & 96$\%$ & 91$\%$ & 79$\%$ \\
S4 & 77$\%$ & 97$\%$ & 96$\%$ & 90$\%$ & 79$\%$ \\
S5 & 73$\%$ & 97$\%$ & 96$\%$ & 89$\%$ & 76$\%$ \\
\hline                                   
\end{tabular}
\end{table}

We must now estimate how many C/2016 R2-like comets could be captured by the Oort Cloud, so as to then evolve dynamically over the next 4 Gyr and return to visit the inner Solar System on C/2016 R2-like orbits. While it is tempting to say that the cometesimals lost from our simulation were ejected from our Solar System, a further investigation of the orbital elements at the moment they were removed from the simulation is required in order to estimate their capture rate by the Oort Cloud. This rate is poorly constrained as of yet as this would depend greatly on the timeline of evolution coinciding with our Sun's ejection from its parent cluster. Further numerical simulations are required in order to investigate the behavior of these comets beyond the 10000 au cutoff, although the effects of Galactic tides 4 Gyr ago are still unknown. Nevertheless, we can make a safe estimate of which comets are bound to the Solar System from the energy $z$ of the cometesimals at the moment they are lost:

\begin{equation}
    z = i_\alpha \frac{GM_\Sun}{2a},
\end{equation}

\noindent where $i_\alpha$ is 1, 0, and -1 for e>1, e=1, and e<1, respectively, $M_\Sun$ the mass of the Sun, and $G$ the gravitational constant. We consider an object captured by the Oort Cloud if the final semi-major axis is $a>10000$ au and its final energy is $z/GM_\Sun \leq 0.00005$ at the moment it is lost from our simulation. Otherwise, we consider it has a truly hyperbolic orbit and is seen as detached from the Solar System.

\begin{figure}
   \centering
   \includegraphics[width=\linewidth]{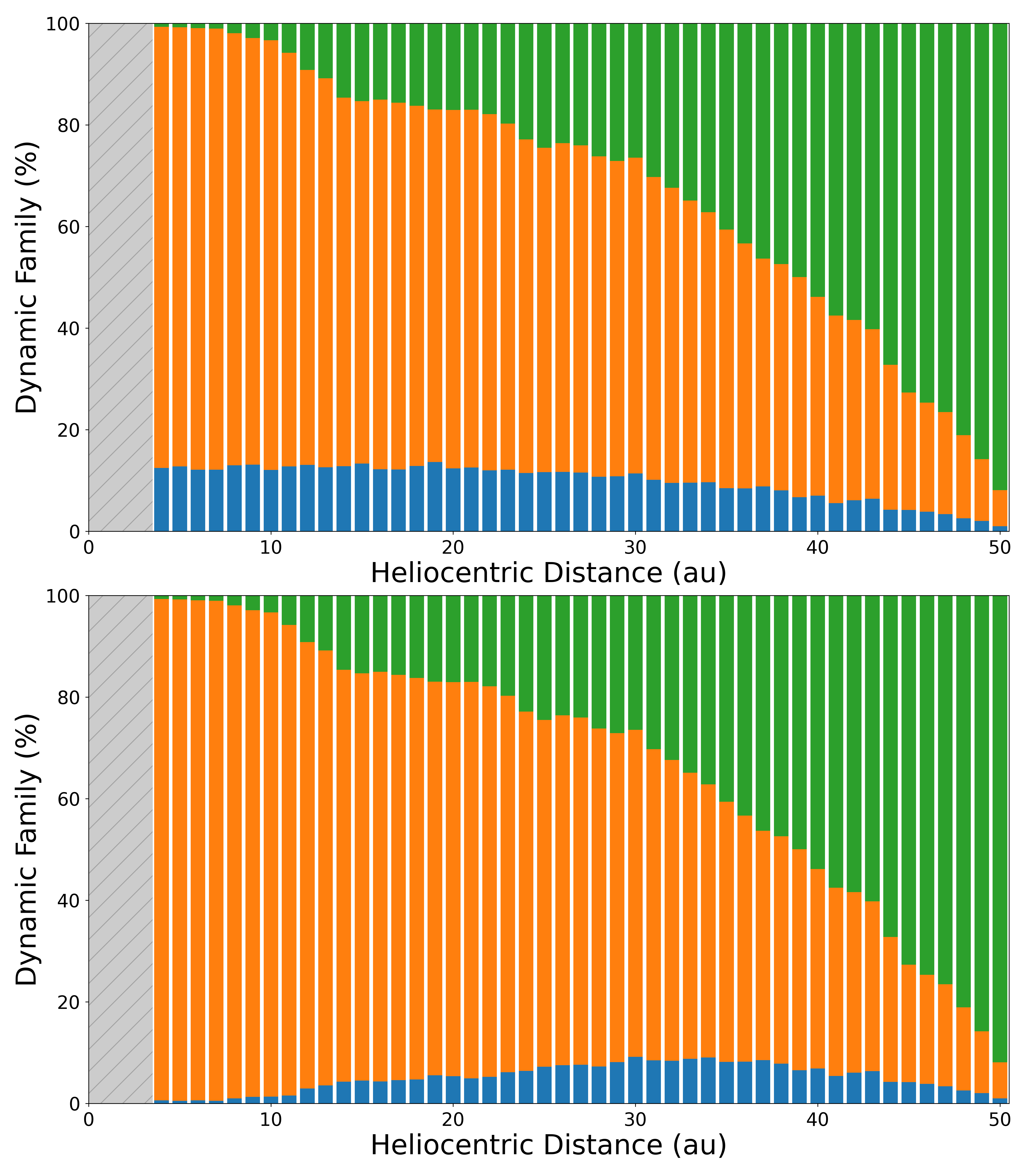}
   \caption{Probability of the fates of cometesimals from each formation zone at the end of 100~Myr  if the sun has left its parent cluster (upper pannel) and if the Sun is still in its parent cluster for the first 10~Myr of our simulation (lower pannel). In orange we see the cometesimals ejected from the Solar System, in blue those captured by the Oort Cloud, and in green those remaining in the simulation.  If the sun has left its parent cluster before the beginning of the simulation, each region <~35~au would contribute $\sim$ 11\% of its clones to the Oort Cloud. If the Sun is still in its parent cluster for the first 10~Myr of the simulation, only 1\% of clones formed between 8 and 11 au are likely to have reached the Oort Cloud, representing 4\% of all the cometesimals potentially captured.}
   \label{fig:Families}
\end{figure}

Under these conditions, 11\% of all our clones have potentially reached the Oort Cloud. The distribution of contributions to the Oort Cloud from each formation zone can be seen in Fig. \ref{fig:Families}. Of the cometesimals formed between 8 and 11 au, 13\% may have reached the Oort Cloud, which represents 12\% of all the cometesimals potentially captured. The 20-30 au formation zone contributes 12\% (making up 22\% of the total number of those captured); the 30-40 au formation zone contributes 9\% (making up 14\% of the total number of those captured); and the 40-50 au formation zone contributes only 5\% (making up 6\% of the total number of those captured). If this estimate is accurate, we should have far more CO-rich comets in the Oort Cloud, between 10\% and 20\% of all long-period comets we observe today. However, this is not the case.

As the timeline of evolution coinciding with our Sun's ejection from its parent cluster is still poorly constrained, we re-estimate our capture conditions. \citet{Zwart2021} estimate that, so long as the Sun is a cluster member, clones with an eccentricity of $e > 0.98$ and a semi-major axis of $a > 2400$ au would be vulnerable to being stripped by the cluster potential or by passing stars. This could not apply to the entire course of our simulation as the Oort Cloud would be unable to form if this were the case. If we align ourselves with \citet{Zwart2021} and consider that our Sun is still within its parent cluster for the first 10~Myr of our simulation, we consider a clone captured if it fulfills the first criteria (final $a>10000$ au final $z/GM_\Sun \leq 0.00005$) along with a new criterion, $e<0.98$ if this event occurs within the first 10~Myr of the simulation. We have drastically different results, as seen in Fig. \ref{fig:Families}. Under these conditions, of the cometesimals formed between 8 and 11 au, only 1\% may have reached the Oort Cloud, which represents 4\% of all the cometesimals potentially captured. The 20-30 au formation zone contributes 7\% (making up 32\% of the total number of those captured); the 30-40 au formation zone contributes 8\% (making up 33\% of the total number of those captured); and the 40-50 au formation zone contributes only 5\% (making up 15\% of the total number of those captured). These results are coherent with our current understanding of the chronology of Oort Cloud formation \citep{Zwart2021}, whereby it is estimated that the bulk or 70\% of the Oort Cloud material originates from the 15-40 au region, and are near what we observe today.

\section{Conclusions}\label{sec:con}

We find that the majority of objects formed between Saturn and the N$_2$ ice line are ejected early and rapidly in the simulation, meaning that even by the time the Jumping Neptune scenario occurs, the clones are already lost. This could explain the lack of N$_2$-rich, CO-rich, and H$_2$O-depleted comets: these were formed in a very narrow region, and that region empties rapidly because of the influence of giant planets. \citet{Zwart2021} call this procession of comet ejections from the 5-11 au zone the `Conveyor Belt', which aptly describes the phenomenon we see here. Objects formed in this region would be ejected early from the Solar System, in less than $\sim$10 Myr, and be unlikely to join the Oort Cloud. Therefore, if N$_2$-rich, CO-rich, and H$_2$O-depleted comets were to have formed under 11 au, $>$90\% of this population would have been ejected from the Solar System without having been captured by the Oort Cloud. As the \citet{Price2021} formation zone for CO-enrichment is tens of astronomical units in width, suggesting that nearly half of all observable comets (~40\%) will be CO-rich, the Oort Cloud should be full of comets of this type. As this is not the case, then we should rule out this model in favor of that of \citet{Mousis2021}. It should also be noted that if this mechanism is indeed the one by which the comets formed and were subsequently captured, then C/2016 R2-like comets may be some of the first long-period comets to have formed, and the earliest to reach the Oort Cloud. There, objects with a nucleus larger than 5 km could survive thousands of orbits, and hypervolatile loss upon possible perihelion passage. This would indicate that C/2016 R2 represents one of the first Oort Cloud objects, which could provide a direct measurement of CO/N$_2$/CH$_4$ ratios in the PSN \citep{Steckloff2021,Davidsson2021,Prialnik2021,Lisse2022}, and would explain why so few N$_2$-rich, CO-rich, and H$_2$O-depleted comets have been observed today.

We must also consider the possibility that many of the cometesimals remaining may have lost their bulk hypervolatile species in the billions of years since their formation, or even within the time frame of our simulation. Pure hypervolatile ices are only stable on gigayear timescales beyond a heliocentric distance of 100 au \citep{Lisse2021}. If this ejection period were to take place at the same time as the sublimative period of the Edgeworth–Kuiper Belt \citep{Lisse2021, Steckloff2021}, then they would only have $\sim$20 Myr to be placed on a trajectory toward the Oort Cloud before they lose their mostly hypervolatile ices to vacuum via insolation heating \citep{Lisse2022}. When looking at both the sublimation chronology and the Oort Cloud formation chronology together, we have a small window of only $\sim$10 Myr in which an object could be ejected from the giant planet region and inserted into the Oort Cloud. Our results are in line with the hypothesis of \citet{Lisse2022},  which states that interstellar object 2I/Borisov was ejected early from its parent system. However, these chronologies (the Sun's ejection from its parent cluster; the sublimative period of the Edgeworth–Kuiper Belt; the Jumping Jupiter/Neptune scenario) are still poorly constrained, even more so when considering how and when these timelines overlap. The window for the  formation of C/2016 R2 in the Conveyor Belt region and subsequent capture by the Oort Cloud could be longer or shorter than our 10 Myr estimate.

We understand that a more quantitative simulation should account for the effect of Galactic tides and perturbations from passing stars from the Sun's birth cluster. However, such an endeavor is far beyond the scope of the present work. The mass distribution of the Galaxy and the position of the Sun in it 4 billion years ago is still unknown, as are the initial mass function and orbital distribution of  the Sun's  birth cluster. The dynamics of the Solar System, in particular in its infancy, are inherently chaotic. Therefore, one would need to run a large number of simulations, varying the Galactic potential and the Sun's birth cluster influence, in order to only get an average statistics of possible events, given that it is unclear whether our Solar system is generic or peculiar.

Another possible explanation for the composition  of C/2016 R2 was presented by \citet{Desch2021}, who suggest that C/2016 R2 could be a fragment of a differentiated KBO surface that was created from an impact during the period of energetic impacts during the 2:1 Jupiter:Saturn resonance epoch. If the CO-rich formation zone is further out, as suggested by \citet{Price2021}, then the objects formed there would have longer to form, impact, and to travel to the Oort Cloud. Further studies of the chronology of the formation of these objects and the time frame for the dynamic instability should be explored in order to investigate the likelihood of these scenarios. A geo-chemical study of how these objects form accompanied with a detailed isotopic and chemical analysis of their current composition would also be beneficial. Understanding whether a C/2016 R2-like object can form with its peculiar composition in situ in the disk or the composition arises from the differentiation of a Pluto-like object would shed light on which of these processes is more likely.

This also allows for the existence of possible exotic comets, with peculiar enrichments stemming from unique composition pockets in the disk. Hypothetically, the ice line of each species would create a small enrichment zone, producing small bodies dominated by this species rather than H$_2$O. By examining the ice lines of the volatile molecules, we can estimate the probabilities of finding comets with each composition. 

\begin{acknowledgements}
We acknowledge the region of Bourgogne-Franche-Comté for their funding of the \texttt{DIAZOTE} project and this work. The project leading to this publication has received funding from the Excellence Initiative of Aix-Marseille Université - A*Midex, a French “Investissements d’Avenir programme” AMX-21-IET-018. 
\end{acknowledgements}


\bibliographystyle{aa}
\bibliography{references.bib}

\end{document}